\documentclass[conference,compsoc]{IEEEtran} 
\AtBeginDocument{%
  }

\usepackage{caption}
\usepackage[table]{xcolor}
\usepackage{tikz}
\usepackage{amsmath,amsfonts}
\usepackage{makecell}
\usepackage{xspace}
\usepackage{booktabs} 
\usepackage{graphicx}
\usepackage{tabularx}
\usepackage{float}
\usepackage{algpseudocode}
\usepackage{subfigure}
\usepackage{makecell} 
\usepackage{multirow}
\usepackage{enumitem}
\usepackage{xurl}

\graphicspath{ {./images/} }
\usepackage{numprint}

\usepackage{etoolbox}
\usepackage{pgfgantt}
\usepackage{rotating}
\usepackage{tabularray}
\usepackage{hyperref}
\usepackage{cleveref}
\usepackage{comment}
\usepackage{algorithm}
\usepackage{algpseudocode}
\usepackage{adjustbox}
\usepackage{colortbl}
\usepackage{xcolor}
\usepackage{inconsolata}

\usepackage{listings}
\usepackage{array}
\definecolor{lightgray}{gray}{0.9}
\usepackage[table]{xcolor}
\usepackage{tcolorbox}
\tcbuselibrary{listings,breakable}
\definecolor{benignblue}{rgb}{0.88, 0.93, 1.0} 
\definecolor{sensitiveorange}{rgb}{1.0, 0.82, 0.72}

\definecolor{codeblue}{rgb}{0.1,0.1,0.7}
\definecolor{codegreen}{rgb}{0.0,0.5,0.0}
\definecolor{codegray}{rgb}{0.4,0.4,0.4}

\newcommand{\kw}[1]{\textcolor{codeblue}{\textbf{#1}}}
\newcommand{\cm}[1]{\textcolor{codegreen}{\textit{#1}}}
\usepackage{pifont}
\newcommand{\cmark}{\textcolor{green!60!black}{\ding{51}}}
\newcommand{\xmark}{\textcolor{red!75!black}{\ding{55}}}

\usepackage[table]{xcolor}

\definecolor{CorrectGreen}{RGB}{34,139,34}
\definecolor{PartialGreen}{RGB}{128,160,60}
\definecolor{FailRed}{RGB}{190,80,80}

\newcommand{\cvec}[1]{\textcolor{CorrectGreen}{#1}}
\newcommand{\cvep}[1]{\textcolor{PartialGreen}{#1}}
\newcommand{\cvef}[1]{\textcolor{FailRed}{#1}}

\newcommand{\algcomment}[1]{\hfill{\color{blue}\footnotesize // #1}}

\definecolor{darkgreen}{rgb}{0.0, 0.5, 0.0}

\newcommand{\etal}{\emph{et~al.}\xspace}
\newcommand{\ournameNoSpace}{\mbox{RAVEN}}
\newcommand{\ourname}{\ournameNoSpace\xspace}

\newcommand{\sys}{\ourname}

\newcommand{\sect}{Sect.~}
\newcommand{\fig}{Fig.~}
\newcommand{\tab}{Tab.~}
\newcommand{\hsect}[1]{\hyperref[#1]{\sect\ref{#1}}}
\newcommand{\hfig}[1]{\hyperref[#1]{\fig\ref{#1}}}
\newcommand{\htab}[1]{\hyperref[#1]{\tab\ref{#1}}}

\usepackage{tikz}

\newcommand{\CHone}{\hyperref[ch:challenge1]{CH1 (\CRSS-Definition)}}
\newcommand{\CHtwo}{\hyperref[ch:challenge2]{CH2 (Dataset Constraints)}}
\newcommand{\CHthree}{\hyperref[ch:challenge3]{CH3 (Long-Range Semantics)}}
\newcommand{\CHfour}{\hyperref[ch:challenge4]{CH4 (Scaling Domain-Specific LLMs)}}

\IEEEoverridecommandlockouts
\makeatletter\def\@IEEEpubidpullup{6.5\baselineskip}\makeatother

\begin{document}

\date{}

\title{RAVEN: Agentic RAG for Automated Vulnerability Repair}

{\author
{\IEEEauthorblockN{Varun Gadey\IEEEauthorrefmark{1},
		Zijie Liu\IEEEauthorrefmark{1},
		Alexandra Dmitrienko\IEEEauthorrefmark{1}}
		\IEEEauthorblockA{\IEEEauthorrefmark{1}University of Duisburg-Essen, Germany\\
		}

}}

\maketitle
  
\begin{abstract}
Automated vulnerability repair has emerged as a promising direction to mitigate the growing number of software vulnerabilities. 
Recent advances in Large Language Models (LLMs) have further accelerated research in automated repair. 
However, existing frameworks remain largely restricted to memory-related vulnerabilities and locally repairable vulnerability settings, leaving generalization to unseen vulnerability types underexplored. 
Their evaluations are often limited to a single programming language, and largely rely on proprietary models.
In this paper, we propose \sys{}, a scalable, efficient and autonomous framework that integrates an agentic retrieval-augmented generation (RAG) pipeline with controlled iterative repair in a unified framework. 
The framework utilizes open-source LLMs in a fully locally deployable setting with limited GPU requirements, while building a multi-faceted retrieval pipeline to retrieve historically relevant vulnerability fixes and guide the patch generation. 
In addition, \sys{} introduces a dedicated Curator Agent that retrieves cross-file dependencies from the target repository, to fix complex vulnerabilities that cannot be addressed using local vulnerable code alone. 
We evaluate \sys{} on 160 real-world CVE vulnerabilities across diverse vulnerability types, two programming languages, unseen CWE categories, and out-of-distribution settings. \sys{} achieves an overall repair success rate of 83.13\%, outperforming all existing state-of-the-art repair frameworks, while also demonstrating strong  generalization capabilities and maintaining the repair cost negligible. 

\end{abstract}

\section{Introduction}
\label{sec:introduction}

Software vulnerabilities remain one of the most exploited entry points for modern cyberattacks, while the volume of disclosed vulnerabilities continues to grow rapidly. In 2025, more than 48K CVEs were published, averaging around 131 disclosures per day~\cite{Vul_Stats_2026}. 
At the same time, remediation remains slow in practice, with critical application vulnerabilities reported in 2025 taking an average of 74.3 days to 
This growing scale and diversity makes manual vulnerability repair increasingly difficult, as developers must analyze program behavior, identify root causes, and fix vulnerabilities without introducing new defects. Consequently, automated vulnerability repair has become essential for producing timely and reliable fixes for both existing and emerging vulnerability threats.

A wide range of automated vulnerability repair (AVR) frameworks have been proposed over the past decade. 
\textit{Search-based} methods such as CapGen~\cite{CAPGEN}, sharpFix~\cite{sharpFix} explore predefined mutation spaces guided by test cases, but often suffer from search space explosion and limited coverage when correct patches are absent. \textit{Semantic-based} approaches, including SOSRepair~\cite{SOSrepair} and Maple~\cite{maple}, leverage symbolic execution and program constraints to guide repair, improving efficiency but restricting adaptability to complex or diverse vulnerabilities. 

\textit{Template-based} approaches, such as PAR~\cite{PAR}, kPAR~\cite{KPAR}, and AVATAR~\cite{AVATAR}, construct patches using predefined fix patterns.
While effective for recurring defects, they remain bounded by the availability and coverage of templates, as well as fault localization accuracy, hindering applicability to diverse and unseen vulnerabilities.

\textit{Learning-based} approaches model program repair as a data-driven task from buggy to fixed code using neural models. Techniques such as Sequencer~\cite{sequence}, DeepRepair~\cite{deeprepair}, and CoCoNuT~\cite{10.1145/3395363.3397369} learn code patterns to improve patch generation.
Despite these advances, their effectiveness remains constrained by training data quality and distribution, often limiting generalization to complex and context-dependent vulnerabilities. 

More recently, \textit{LLM-based} approaches~\cite{vrepair, seqtrans, Vulrepair, Vulmaster, zeroshotrepair, rapgen, san2patch, patchagent, appatch} have emerged as a promising direction by leveraging large code models to reason over vulnerable code and to generate patches.
Existing works include training or fine-tuning based repair frameworks such as VRepair~\cite{vrepair}, SeqTrans~\cite{seqtrans}, VulRepair~\cite{Vulrepair}, and VulMaster~\cite{Vulmaster}, retrieval-augmented generation for functionality bugs such as RAP-Gen~\cite{rapgen}, and zero-shot or prompt-based repair approaches such as Pearce \etal~\cite{zeroshotrepair}. 
Among recent state-of-the-art frameworks, PatchAgent~\cite{patchagent} introduces an autonomous agent-based framework that combines fault localization, patch generation, and validation. 
APPATCH~\cite{appatch} proposes adaptive prompting and vulnerability semantics reasoning for patch generation. 
SAN2PATCH~\cite{san2patch} leverages sanitizer logs with multi-stage LLM reasoning to decompose vulnerability repair into structured subtasks. 

Despite these advances, existing LLM-based frameworks remain limited in several important aspects. 
They are often restricted to specific Common Weakness Enumerations (CWEs) i.e. categories; and are commonly evaluated only within a single programming language; and mostly target vulnerabilities whose repair can be inferred from a single line, function, or local code region. 
Moreover, their generalization capability to unseen vulnerability types remains underexplored, and dedicated mechanisms for controlling the probabilistic nature of LLMs are largely ignored. 
Collectively, these limitations reveal \textbf{open challenges} in automated vulnerability repair, which are discussed in detail below.

\textbf{\textit{Challenge} \textcircled{1}} Many real-world vulnerabilities (CVEs) can span across multiple source files and cannot be repaired using only the local code snippet. 
For complex families of vulnerability types such as access control, input handling, and injection-based, the root cause may depend on cross-file dependencies, global context or semantic and data-flow interactions across the repository. 
However, recent repair frameworks~\cite{appatch, patchagent, san2patch, ExtractFix, Vulmaster, vrepair} largely focus on memory-safety and arithmetic-related families of vulnerability types, where the repair is often very localized.

\textbf{\textit{Challenge} \textcircled{2}} Due to the probabilistic nature of LLMs, generated patches may unintentionally introduce new vulnerabilities or provide only partial fixes to existing ones.  
Moreover, all the recent frameworks~\cite{san2patch, rapgen, zeroshotrepair, appatch, patchagent, ExtractFix, Vulmaster, vrepair} primarily focus only on one-shot automated repair. 
Dedicated mechanisms for systematically scrutinizing, reviewing, and refining generated patches through iterative feedback remain largely unexplored.

\textbf{\textit{Challenge} \textcircled{3}} Most of the LLM-based repair frameworks~\cite{patchagent, appatch} heavily rely on large proprietary models from providers such as OpenAI and Anthropic, leading to high token cost, longer repair time, and limited deployability.
For instance, PatchAgent~\cite{patchagent} reports that running only 75 repair cases cost over \$1500, highlighting the financial limitations of such approaches. 
Thus, efficient, low-cost, and locally deployable repair framework remains an important open challenge.

\textbf{\textit{Challenge} \textcircled{4}} Finally, LLM-based repair frameworks remain highly susceptible to hallucinations, inconsistent patch generation, and non-deterministic behaviors. 
Although recent works such as APPATCH~\cite{appatch} acknowledge this issue through post-generation validation strategies, dedicated mechanisms to proactively reduce hallucinations during patch generation itself remain largely unexplored.

To address these challenges, we propose \sys{}, a scalable and autonomous framework for automated vulnerability repair designed to operate across diverse vulnerability types  and generalize to previously unseen vulnerabilities. 
\sys{} is the first repair framework to integrate an agentic retrieval-augmented generation (RAG) pipeline with historical fix retrieval, context dependency retrieval, and controlled iterative patch refinement in a unified framework. 

To support practical and cost-effective deployment, \sys{} utilizes recent open-source medium-sized LLMs, including Gemma-4-26B~\cite{gemma4_2026}, and Nemotron-3-30B~\cite{nemotron_30b},
allowing locally deployable vulnerability repair without reliance on proprietary cloud-based models.
With \sys{}, developers can efficiently perform automated vulnerability repair across diverse software projects using locally deployable open-source LLMs, reducing significantly repair cost and dependency on proprietary services. 

\textbf{Contributions} In particular, we make the following contributions:

\begin{itemize}
    \item We propose \sys{}, a first-of-its-kind automated vulnerability repair framework designed that uniquely combines the ability to repair complex real-world vulnerabilities across \textbf{10} diverse CWE types and programming languages, including Java and C.

    \item We design an agentic RAG framework that performs multi-faceted retrieval using lexical, syntactic, and semantic vulnerability cues to retrieve relevant historical fixes. In addition, \sys{} introduces a dedicated Curator Agent that scans the target repository to extract root-cause information, and cross-file dependencies guiding patch generation beyond local vulnerable snippets.

    \item We introduce an automated iterative repair mechanism that incorporates a dedicated Patch Reviewer module and static analysis feedback to progressively refine generated patches and improve repair quality.

    \item We evaluate \sys{} on \textbf{160} real-world CVE vulnerabilities across diverse experimental setups, including multiple vulnerability types, programming languages, unseen CWE categories, and out-of-distribution datasets. \sys{} achieves an overall repair success rate of \textbf{83.13\%}, outperforming existing state-of-the-art repair tools.

    \item We demonstrate the generalizability and efficiency of \sys{} through cross-language evaluation (\textbf{86.21\%} success rate), unseen vulnerability-type (CWEs) evaluation (\textbf{87.50\%} success rate), and local deployment using open-source LLMs with negligible token cost. 
\end{itemize}

Overall, this work addresses key limitations in existing automated vulnerability repair by proposing a unified agentic RAG-based framework that integrates historical fix and dependency retrieval, iterative patch refinement, and independent validation to enable scalable, low-cost, and generalizable repair of real-world CVEs.
To the best of our knowledge, no existing repair framework explicitly targets Java-based CWEs spanning access control, SQL injection, code injection, and memory-related vulnerabilities. 

\noindent\textbf{Outline.} Appendix~\ref{sec:background} provides background on Retrieval Augmented Generation (RAG).  Section~\ref{sec:motivation} highlights the complexity of fixing vulnerabilities autonomously. In Section~\ref{sec: approach}, we present our \sys approach. Section~\ref{sec:Evaluation} presents the evaluation results,  showcasing \sys's effectiveness across various experimental settings. Section~\ref{sec:related_work} reviews the related literature. Finally, Section~\ref{sec:conclusion} concludes the paper.

\section{Motivation}
\label{sec:motivation}
 
\begin{figure}[t]
\centering
\footnotesize
\begin{tabularx}{\columnwidth}{r X}
1  & \texttt{\kw{this}.modelBridge.setContextUserReference(} \\
   & \hspace*{1em}\texttt{\kw{this}.request.getUserReference());} \\
2  & \kw{try} \texttt{\{} \\
3  & \hspace*{1em}\texttt{\kw{this}.progressManager.startStep(\kw{this});} \\
4  & \hspace*{1em}\texttt{moveAttachment(source, destination,} \\
   & \hspace*{2em}\texttt{autoRedirect, wiki);} \hspace*{1em}\cm{// vulnerable call} \\
5  & \hspace*{1em}\texttt{\kw{this}.progressManager.endStep(\kw{this});} \\
6  &  \\
7  & \hspace*{1em}\texttt{\kw{this}.progressManager.startStep(\kw{this});} \\
8  & \hspace*{1em}\texttt{\kw{this}.observationManager.notify(} \\
   & \hspace*{2em}\texttt{\kw{new} AttachmentMovedEvent(} \\
   & \hspace*{3em}\texttt{(AttachmentReference) source, destination),} \\
   & \hspace*{2em}\texttt{\kw{this}, \kw{this}.request);} \\
9  & \hspace*{1em}\texttt{\kw{this}.progressManager.endStep(\kw{this});} \\
10 & \texttt{\}} \\
\end{tabularx}
\vspace{-1ex}
\caption{CWE-862 Missing Authorization}
\label{fig:cve2023_code_before}
\end{figure}
In this section, we demonstrate the significance of \sys by examining a real-world vulnerability, \textbf{CVE-2023-37910}\cite{CVE2023_37910}, identified in the XWiki platform\cite{xwiki_platform} within the file \texttt{MoveAttachmentJob.java}. 
This vulnerability corresponds to \textbf{CWE-862 Missing Authorization}\cite{CWE862}, where a security-sensitive operation is executed without verifying whether the requesting user has sufficient access rights.
This class implements a refactoring job responsible for moving attachments across documents in the repository. The vulnerability stems from the method \texttt{protected void process(EntityReference source)}, where the attachment move operation is executed, followed by the triggering of notification events.
As shown in \Cref{fig:cve2023_code_before}, the critical issue occurs at line $4$, where \texttt{moveAttachment(...)} is directly invoked without any prior authorization check on the requesting user. 
The surrounding lines primarily manage execution flow, including setting the user context (lines 1–2), tracking progress (lines 3, 5, 7, 9), but do not perform access control checks before the sensitive operation is executed.

Exploitation of this vulnerability allows an unauthorized user to move attachments across documents without possessing the required access rights, potentially leading to unauthorized data manipulation, or information disclosure  within the repository. 
Fixing this issue is non-trivial and particularly challenging for automated vulnerability repair tools, as the vulnerability arises from a missing security precondition rather than an explicit syntactic or semantic error. 
The repair requires coordinating multiple code-level changes, and integrating the authorization logic at the correct point in the existing control flow.
Moreover, this authorization logic is not fully apparent from the local code alone, meaning that a naive repair strategy focused solely on local context may fail to identify the missing precondition or introduce ad hoc checks that are inconsistent with the system’s design.

\section{RAVEN Design}
\label{sec: approach}

In this section, we present Agentic \underline{\textbf{R}}etrieval \underline{\textbf{A}}ugmented Generation for Automated \underline{\textbf{V}}ulnerability R\underline{\textbf{e}}pair (\textbf{RAVEN}), a scalable and efficient framework designed to autonomously repair real-world software vulnerabilities. RAVEN introduces retrieval-augmented generation (RAG) into the automated vulnerability repair pipeline, enabling the model to leverage external security knowledge during patch generation. Building on this foundation, RAVEN further incorporates an agentic RAG methodology, where the retrieval process is actively guided by lexical, semantic, and syntactic vulnerability patterns derived from publicly disclosed vulnerabilities (e.g., CVEs), while agent-based reasoning dynamically identifies and retrieves relevant contextual information from the project repository. This enables the framework to capture cross-file and global code dependencies automatically and perform more context-aware repair beyond localized code regions.

To further improve repair quality, RAVEN employs an iterative refinement mechanism in which generated patches are progressively evaluated and revised using feedback from a static analysis tool and a patch review module. This design aims to support generalization across programming languages and diverse vulnerability types, addressing limitations observed in recent state-of-the-art repair approaches~\cite{patchagent, appatch, san2patch}. \sys uses recent medium-sized LLMs, such as Gemma-4-26B~\cite{gemma4_2026} and Nemotron-3-30B~\cite{nemotron_30b}, as core components for patch generation and evaluation. These models offer a practical balance between effectiveness and efficiency and can be deployed locally, allowing \sys to maintain a lightweight and cost-effective design without reliance on external services.

\begin{figure}[t]
    \centering
    \includegraphics[width=0.85\columnwidth]{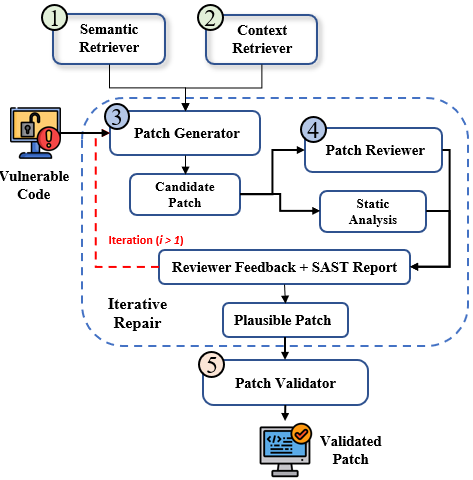} 
    \caption{\sys Framework Overview;}
    \label{fig:cal_phases_overview}
\end{figure}

\begin{figure*}[t]
    \centering
    \includegraphics[width=1\textwidth]{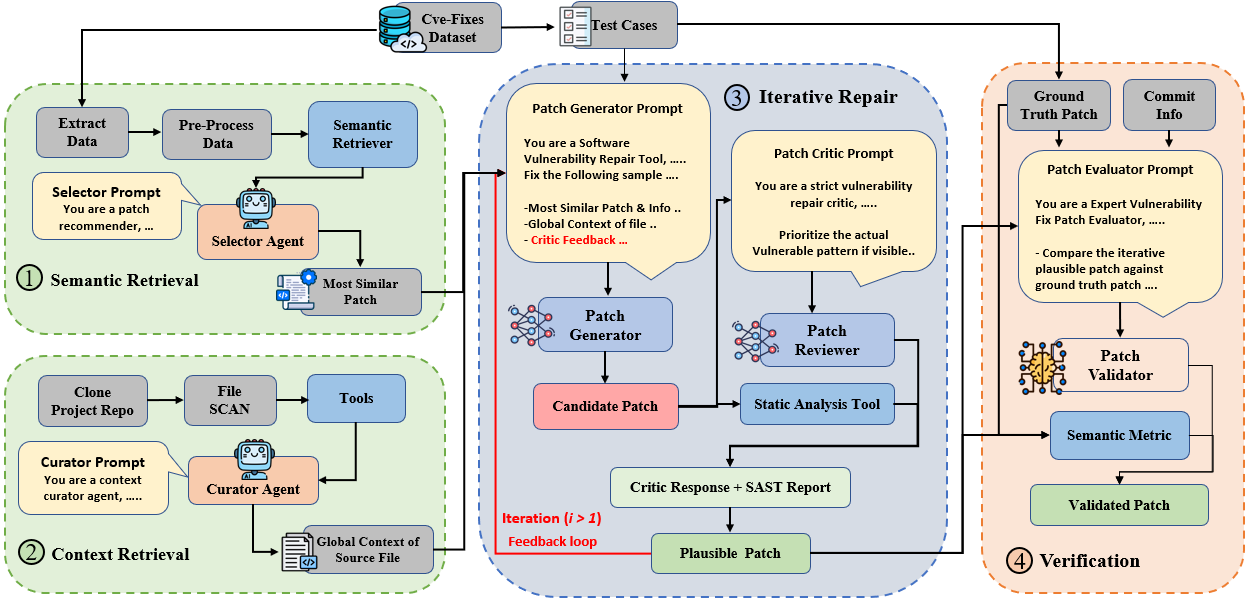} 
    \caption{\sys's Repair Workflow with its four phases. \textcircled{1}\textsf{Similar Retrieval} and \textcircled{2} \textsf{Context Retrieval} extracts and provides various useful information to the Patch Generator. \textcircled{3} \textsf{Iterative repair} brings out the best plausible patch for the vulnerability through numerous iterations between generator and reviewer. \textcircled{4} \textsf{Verification} where the plausible patch is validated against the ground truth patch; Color schema: Gray $\rightarrow$ Data, orange $\rightarrow$ Agents, Yellow $\rightarrow$ prompts, Blue $\rightarrow$ Tools, Red $\rightarrow$ Vulnerable, Green $\rightarrow$ Fix} 
    \label{fig:cal_inference}
\end{figure*}

\subsection{Framework Overview}
\label{sec:framework_overview}

\Cref{fig:cal_phases_overview} provides an overview of the \sys{} framework, illustrating its core modules and their interactions.
\sys comprises five core modules that are \textcircled{1} \textsf{Semantic Retriever}, \textcircled{2} \textsf{Context Retriever}, \textcircled{3} \textsf{Patch Generator}, \textcircled{4}~\textsf{Patch Reviewer}, and \textcircled{5} \textsf{Patch Validator}. 
Each module has its own dedicated pipeline and is designed to meet a specific objective. For instance, \textsf{Sematic Retrieval} module is responsible for retrieving highly-relevant and semantically similar fix patches, while \textsf{Context Retrieval} is designed to address the complex vulnerabilities by providing cross file dependencies and global context of the vulnerable code.  The outputs of both these modules are serving as inputs to \textsf{Patch Generator} module. 
One-shot vulnerability repair rarely succeeds across various vulnerability types. Therefore, \textsf{Patch Generator} and \textsf{Patch Reviewer} modules are involved into an automated iterative repair process, where \textsf{Patch Generator} produces candidate patch, which is reviewed by \textsf{Patch Reviewer} and a static analysis tool. 
Lastly, the best patch among all the patches generated through the iterations is served as an input to \textsf{Patch Validator} module. 
\textsf{Patch Validator} uses a dedicated LLM configuration and standard code semantic metrics to independently assess the quality of the final patch before it is reported as validated. 

\subsection{Repair Work Flow} 
\label{sec:detailed_design}

\Cref{fig:cal_inference} illustrates the end-to-end repair workflow of \sys{}.
There are four key work phases in the framework \textcircled{1}~\textsf{Semantic Retrieval},  \textcircled{2}~\textsf{Context Retrieval}, \textcircled{3}~\textsf{Iterative Repair} and \textcircled{4}~\textsf{Verification}. 

\begin{figure}[t]
    \centering
    \includegraphics[width=0.88\columnwidth]{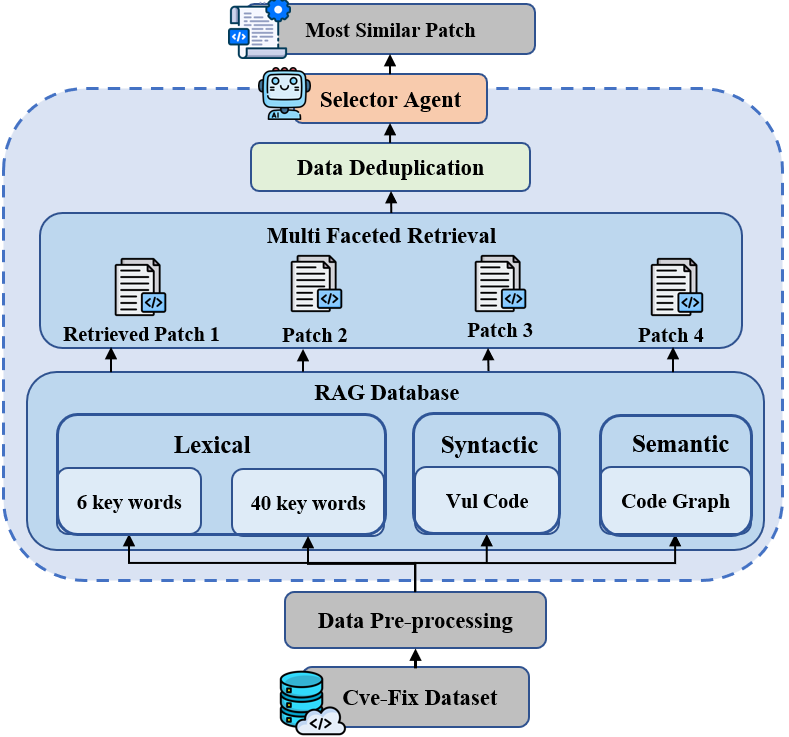} 
    \caption{Semantic Retriever Module}
    \label{fig:semantic_retriever}
\end{figure}

\subsubsection{Semantic Retrieval} 
\label{sec:semantic_retrieval}

The \textsf{Semantic Retriever} module mitigates spurious pattern learning and hallucinations in the \textsf{Patch Generator} by retrieving highly relevant and semantically similar historical vulnerability fixes to guide patch generation. To improve retrieval robustness, the module employs multiple specialized retrievers, each designed to capture different vulnerability-related patterns in the vulnerable code, enabling a multi-faceted retrieval process.
The goal of the \textsf{Semantic Retriever} is to dynamically identify the most relevant historical fix patches corresponding to the same or closely related vulnerability types. These retrieved patches are augmented into the prompt template of the \textsf{Patch Generator}, directing and constraining the repair process toward more accurate and consistent patch generation.

To support this retrieval process, we construct a dedicated RAG database as shown in~\Cref{fig:semantic_retriever}) from a comprehensive real-world Common Vulnerabilities and Exposures (CVE) dataset, namely the CVEFixes\cite{CVEFixes} dataset. The database contains diverse vulnerability instances, their corresponding fix patches, and associated metadata. Specifically, we curate a broad set of CVEs covering the targeted CWEs, enabling retrieval across a wide range of vulnerability patterns. As new vulnerability types and recently disclosed CVEs become available, the database can be continuously extended, making \sys adaptable and scalable.
Retrieving the most relevant vulnerability patch, however, remains challenging. Including multiple retrieved patches in the prompt may introduce ambiguity, potentially confusing the LLM and diverting the repair process. Likewise, excessive contextual information can act as noise within the model’s limited context window. Effective retrieval and careful prompt context management are therefore critical for reliable repair.

To address these challenges, we introduce a multi-faceted retrieval strategy that retrieves candidate patches based on four complementary vulnerability patterns. Each retriever operates independently on a specific representation and recommends only the single most relevant patch according to its retrieval criterion. Because retrieval is performed across distinct lexical, syntactic, and semantic perspectives, the resulting recommendations are often diverse and complementary.
To enable this process, we perform a dedicated data preprocessing step that transforms historical vulnerability and fix data into multiple (particularly, four) representations capturing lexical, syntactic, and semantic repair cues, as illustrated in \Cref{fig:semantic_retriever}. These representations form the foundation of the retrievers within the \textsf{Semantic Retriever} module. A \textsf{Selector Agent} then evaluates the retrieved candidates and identifies the single most suitable patch to guide repair according to predefined selection criteria.

\noindent \textbf{Preprocessed Representations} Below, we describe the four distinct preprocessed representations used in the \textsf{Semantic Retriever}.

\noindent\emph{6 Keyword:} 
A concise six-keyword representation is constructed by extracting the most salient lexical cues from the CVE description and the corresponding vulnerable code context. This representation is particularly effective for frequently occurring vulnerabilities within the same CWE, where a small set of discriminative lexical signals is often sufficient for accurate retrieval.

\noindent\emph{40 Keyword:} 
A richer forty-keyword representation is constructed using an expanded set of lexical cues derived from the CVE description, code context, and associated metadata. The choice of forty keywords is based on empirical evaluation, balancing broader contextual coverage with the need to minimize retrieval noise.

\noindent\emph{Vulnerable Code:} 
The exact vulnerable code snippet is extracted and stored as a separate representation, preserving the syntactic structure of the vulnerability. This pattern leverages syntax-level cues, which can be particularly effective for vulnerability types where structural code patterns provide strong retrieval signals.

\noindent\emph{Code Property Graphs (CPGs):} 
Code Property Graphs are generated using the Joern tool~\cite{joern}, providing a unified representation of the vulnerable code’s syntax, control flow, and data dependencies. This representation enables deeper semantic retrieval by capturing structural and behavioral similarities across vulnerability instances.


\noindent\textbf{Retrieval Mechanism} 
A sparse retrieval mechanism such as BM25~\cite{bm25} is used to query the RAG database for each preprocessed representation, enabling a multi-faceted retrieval process. Each retriever independently retrieves the most relevant historical vulnerability patch for its respective pattern. A deduplication step removes overlapping results to ensure diversity in the final candidate set. The resulting patches are then passed to the \textsf{Selector Agent} (see \Cref{fig:semantic_retriever}), which evaluates them using predefined criteria and selects the single most relevant patch to guide the \textsf{Patch Generator} module.

\noindent\textbf{Selection Criteria} 
The \textsf{Selector Agent} evaluates candidate patches based on the following key selection criteria: (1)~prioritizing the underlying vulnerability root cause over superficial attributes such as file names, function names, or CVE identifiers; (2)~favoring patches that exhibit the most similar repair strategy and security logic, including bounds validation, length checks, null handling, and state consistency; (3)~avoiding reliance on superficial textual overlap and instead emphasizing semantic equivalence in the applied fix; (4)~resolving inconsistencies between the CWE description and the vulnerable code by prioritizing the actual code pattern and its implied root cause; and (5)~treating patches that replicate logic already present in the vulnerable code with caution to account for dataset noise or previously patched code fragments.
The prompt template for semantic retrieval is provided in the Appendix~\ref{appendix:templates}.

\subsubsection{Context Retrieval} 
\label{sec:context_retrieval} 

Most of the existing vulnerability repair tools{\cite{appatch, san2patch, patchagent}} address vulnerabilities based on the vulnerable code, logs, metadata, and recommended exemplars, which constrain the tool's access to only the local context of the vulnerable code and which can significantly impact the success of automated repair. 
Moreover, for certain vulnerability types, it is nearly impossible to accurately analyze and fix the vulnerability by considering only the local context of the code. 
Addressing this challenge, \sys{} introduces a dedicated \textsf{Context Retriever} module, which operates in parallel with the \textsf{Semantic Retriever} module to retrieve global context and dependency code snippets from the project repository. Specifically, A dedicated Curator Agent is designed to retrieve these dependencies as shown in~\Cref{fig:cal_inference}. The retrieved context is then provided as an additional input to the Patch Generator module ultimately guiding the repair process. 

\Cref{alg:context_curator} illustrates the detailed workflow of the \textsf{Context Retriever} module. 
The module takes as input the vulnerable code $v$, vulnerability metadata $c$, and target file path $f$ and produces a potential root-cause description, repair constraints, and single or multiple dependency code snippets as output.  
We design and equip the Curator agent with five tool-calling operations, namely \textcircled{1}~\texttt{read\_bundle}, \textcircled{2}~\texttt{search\_repo}, \textcircled{3}~\texttt{find\_symbol}, \textcircled{4}~\texttt{read\_file}, and \textcircled{5}~\texttt{finish\_curating}.
The exploration is performed iteratively over a bounded number of steps (line~4 in \cref{alg:context_curator}), where at each iteration the agent selects a single tool based on the current context and accumulated observations to progressively refine its understanding of the vulnerability.

The Curator Agent begins exploration with an initial \textcircled{1}~\texttt{read\_bundle} call (line~5 in \cref{alg:context_curator}), which provides a structured summary of the target file, including method-level information, related symbols, and caller--callee relationships. 
Subsequently, the Agent utilizes \textcircled{2}~\texttt{search\_repo} (line~6 in \cref{alg:context_curator}) to locate relevant code patterns across the repository using keyword or regex-based search. 
It employs \textcircled{3}~\texttt{find\_symbol} (line~7 in \cref{alg:context_curator}) to identify symbol definitions and trace caller relationships for cross-file dependency analysis. 
Further, \textcircled{4}~\texttt{read\_file} (line~8 in \cref{alg:context_curator}) is used to inspect selected file regions from the repository in detail and extract relevant dependency code snippets. 
Finally, \textcircled{5}~\texttt{finish\_curating} (line~9 in \cref{alg:context_curator}) explicitly terminates the exploration and returns the found dependency code snippets, root cause and a list of repair constraints. 
This Agentic exploration loop continues until a hard limit of $10$ steps is reached. The $10$ step limit is chosen empirically as a balance between collecting sufficient dependency context and avoiding unproductive exploration. A shorter limit may risks missing key dependency context, whereas an unbounded loop would significantly increase runtime cost.

\begin{algorithm}[t]
\caption{Context Retrieval}
\label{alg:context_curator}
\begin{algorithmic}[1]
\Statex \textbf{Input:} metadata $c$, file path $f$, vulnerable code $v$
\Statex \textbf{Output:} Root cause, repair constraints, dependency code

\State Initialize trace $\mathcal{T} \gets \emptyset$
\State Load $v$ from $f$ and metadata $c$
\State Construct prompt with metadata $c$, file $f$, \& trace $\mathcal{T}$

\For{$i = 1$ to $N$}
    \algcomment{Agent selects next tool}

    \If{agent calls \texttt{read\_bundle}}
        \algcomment{Load structured summary of the target file, symbols, callers, callees, tests, build files}
    \ElsIf{agent calls \texttt{search\_repo}}
        \algcomment{Search repo for relevant patterns with literal or regex search}
    \ElsIf{agent calls \texttt{find\_symbol}}
        \algcomment{Retrieve symbol definitions and caller relationships}
    \ElsIf{agent calls \texttt{read\_file}}
        \algcomment{Read selected line ranges from repo files for detailed inspection}
    \ElsIf{agent calls \texttt{finish\_curating}}
        \State \Return root cause, dependency code, \& constraints
    \EndIf

    \State Append tool observation to trace $\mathcal{T}$
\EndFor

\State \Return retrieved context 
\label{alg:context_curator}
\end{algorithmic}
\end{algorithm}

\subsubsection{Iterative Repair} 
\label{sec:iterative_repair} 

Controlling the behavior of the LLM within the \textsf{Patch Generator} is challenging due to the probabilistic nature of language models. To guide patch generation effectively, \sys injects the outputs of the \textsf{Semantic Retriever} and \textsf{Context Retriever} into the prompt template of the \textsf{Patch Generator}. The prompt assigns a specific repair objective, defines explicit rules and constraints, and provides all necessary contextual information in a concise manner. Based on this augmented context, the \textsf{Patch Generator} produces an initial fix patch for the given vulnerability.
To further improve patch quality, \sys employs an iterative feedback loop between the \textsf{Patch Generator} and the \textsf{Patch Reviewer}, as illustrated in the middle section of \Cref{fig:cal_inference}. The \textsf{Patch Reviewer}, instantiated using the same LLM model as \textsf{Patch Generator}, evaluates each generated patch according to predefined repair criteria. In addition, a static application security testing (SAST) tool, Semgrep~\cite{semgrep_github}, is used to analyze the generated patch for potential security issues and rule violations. The reviewer’s structured feedback, together with the SAST report, is then returned to the \textsf{Patch Generator} to guide subsequent repair iterations.
This iterative refinement process enables the framework to progressively improve patch correctness and reduce ineffective or suboptimal repairs.

\noindent\textbf{Patch Generator:} 
The \textsf{Patch Generator} is guided by the following principles: 
\textbf{\textcircled{1}}~prioritizing the concrete vulnerability root cause over superficial similarity to retrieved patches; 
\textbf{\textcircled{2}}~applying the smallest targeted fix necessary to eliminate the unsafe behavior while preserving existing program logic; 
\textbf{\textcircled{3}}~avoiding unnecessary refactoring, renaming, formatting changes, or modifications to unrelated code regions; 
\textbf{\textcircled{4}}~refraining from introducing new helper functions, macros, types, or external dependencies unless clearly implied by the provided context; 
\textbf{\textcircled{5}} reusing existing validation patterns, control flow, and error-handling mechanisms to maintain consistency with the surrounding code; 
\textbf{\textcircled{6}} ensuring syntactic correctness and coherence with the local code structure; 
\textbf{\textcircled{7}}~restricting the output to a single code block containing only the minimal fixed code snippet(s); and 
\textbf{\textcircled{8}}~omitting any additional explanatory text outside the code block. 
Additional language-specific and revision-related requirements are provided in the prompt templates in Appendix~\ref{appendix:templates}.

\noindent\textbf{Patch Reviewer:} \textsf{Patch Reviewer} evaluates each generated patch according to the following criteria: 
\textbf{\textcircled{1}}~disregarding unused helper methods, dead code, comments, logging statements, and formatting-only changes as valid repairs; 
\textbf{\textcircled{2}}~verifying that the introduced security logic is executed on the actual runtime path before or at the vulnerable sink; 
\textbf{\textcircled{3}}~ensuring that, for vulnerabilities involving library misuse, the patch replaces, intercepts, or mitigates the exact affected runtime behavior; 
\textbf{\textcircled{4}}~comparing the current patch against the previous iteration strictly in terms of vulnerability-fixing effectiveness; and 
\textbf{\textcircled{5}}~identifying regressions only when a new patch weakens existing protection, reintroduces the vulnerability, or replaces an effective fix with an ineffective or irrelevant modification.

Controlling the behavior of the LLM within the \textsf{Patch Generator} is challenging due to the probabilistic nature of language models. To guide patch generation effectively, \sys injects the outputs of the \textsf{Semantic Retriever} and \textsf{Context Retriever} into the prompt template of the \textsf{Patch Generator}. The prompt assigns a specific repair objective, defines explicit rules and constraints, and provides all necessary contextual information in a concise manner. Based on this augmented context, the \textsf{Patch Generator} produces an initial fix patch for the given vulnerability.

To further improve patch quality, \sys employs an iterative feedback loop between the \textsf{Patch Generator} and the \textsf{Patch Reviewer}, as illustrated in the middle section of \Cref{fig:cal_inference}. The \textsf{Patch Reviewer}, instantiated using the same LLM model as the generator, evaluates each generated patch according to predefined repair criteria. In addition, Semgrep~\cite{semgrep_github}, is used to analyze the generated patch for potential security issues and rule violations. \textsf{Patch Reviewer}'s structured feedback, together with the SAST report, is then returned to the \textsf{Patch Generator} to guide subsequent repair iterations.

\subsubsection{Verification} 
\label{sec:validator} 
Right side of \Cref{fig:cal_inference} illustrates the final \textsf{Patch Validator} module, which evaluates the most promising patch produced by the iterative repair process. This stage performs multi-faceted validation by combining LLM-based assessment with quantitative semantic similarity analysis.

The generated patch is first compared against the ground truth fix, which is available for each real-world CVE in the evaluation dataset. A dedicated \textsf{Patch Validator}, instantiated using a distinct LLM from the previous modules, acts as an expert judge under a strict evaluation guideline. The validator assesses the generated patch in relation to the ground truth patch and uses commit information as additional input, considering factors such as repair correctness, completeness, alignment with the intended vulnerability resolution, and consistency with expected repair patterns. Based on this assessment, each patch is classified into one of three categories: \textit{Full Fix Patch}, \textit{Partial Fix}, or \textit{No Fix}.

To complement the LLM-based judgment, a CodeBLEU~\cite{codebleu} score is computed between the generated patch and the ground truth fix. Unlike exact string matching, CodeBLEU captures structural and semantic similarity, providing a quantitative measure of repair quality.
Together, the LLM-based evaluation and CodeBLEU scoring provide a robust validation strategy for assessing both functional correctness and semantic alignment with the expected fix.

\noindent\textbf{Summary} Overall, RAVEN is a modular agentic framework that combines multi-faceted retrieval, repository-level context reasoning, iterative patch refinement, and independent validation to enable robust vulnerability repair. It systematically guides LLM-based patch generation toward root-cause fixes while improving reliability through structured feedback and semantic verification.

\section{Evaluation}
\label{sec:Evaluation}

This section evaluates \sys{} across diverse vulnerability types in Java and C, focusing on its ability to generate correct fixes.

\subsection{Targeted Vulnerabilities}

To evaluate the robustness and generalizability of \sys{}, we select a diverse set of real-world vulnerability types based on widely reported CWEs. 
The selected vulnerabilities span multiple categories, including memory safety, injection, access control, and input handling, ensuring broad coverage across different root causes and exploitation patterns. 
This diversity allows us to assess the effectiveness of \sys{} in handling both localized vulnerabilities and those requiring cross-file dependencies and semantic understanding. 
The complete list of targeted CWEs and their categorization is shown in \Cref{tab:target_cwe}. 

\begin{table}[t]
\centering
\renewcommand{\arraystretch}{1.2}
\begin{tabularx}{0.98\columnwidth}{|c|X|c|}
\hline
\textbf{CWE} & \textbf{Description} & \textbf{Cluster} \\
\hline
787\cite{CWE787} & Out-of-bounds Write & Memory \\
125\cite{CWE125} & Out-of-bounds Read & Memory \\
79\cite{CWE79}  & Cross-site Scripting (XSS) & Injection \\
89\cite{CWE89}  & SQL Injection & Injection \\
94\cite{CWE94}  & Code Injection & Injection \\
22\cite{CWE22}  & Path Traversal & Input Handling \\
502\cite{CWE502} & Untrusted Deserialization & Input Handling \\
352\cite{CWE352} & Cross-Site Request Forgery & Access Control \\ 
862\cite{CWE862} & Missing Authorization & Access Control \\
863\cite{CWE863} & Incorrect Authorization & Access Control \\
\hline
\end{tabularx}
\vspace{-0.5em}
\caption{Targeted 10 Java CWE categories}
\label{tab:target_cwe}
\vspace{-0.5em}
\end{table}

\vspace{0.2cm} \noindent\textbf{Datasets:} We leverage publicly available real-world vulnerability datasets to evaluate \sys{}.
Our primary dataset is \textit{CVEFixes}~\cite{CVEFixes}, from which we extract and preprocess vulnerabilities corresponding to $10$ targeted Java-based CWE categories. 
To assess generalization beyond the primary dataset, we additionally evaluate \sys{} on the \textit{Vul4J} dataset~\cite{vul4j2022}, which contains real-world Java vulnerabilities outside our primary CVEFixes~\cite{CVEFixes} dataset and serves as an out-of-distribution benchmark. 
To assess \sys{} on C programs, we further randomly select vulnerabilities corresponding to CWE-125~\cite{CWE125} (\textit{Out-of-bounds Read}) and CWE-863~\cite{CWE863} (\textit{Incorrect Authorization}) from the \textit{CVEFixes}~\cite{CVEFixes}.

Furthermore, we adopt a subset of the \textit{Zero-Day} C language dataset introduced in APPATCH~\cite{appatch}, consisting of vulnerabilities reported after 2024, to evaluate \sys on recently disclosed and diverse vulnerability instances.
This subset includes diverse unseen vulnerability types such as CWE-835~\cite{CWE835} (\textit{Loop with Unreachable Exit Condition}), CWE-828~\cite{CWE828} (\textit{Signal Handler with Non-reentrant Function}), CWE-770~\cite{CWE770} (\textit{Allocation of Resources Without Limits or Throttling}), CWE-476~\cite{CWE476} (\textit{NULL Pointer Dereference}), CWE-416~\cite{CWE416} (\textit{Use After Free}), CWE-281~\cite{CWE281} (\textit{Improper Preservation of Permissions}), and CWE-96~\cite{CWE96} (\textit{Improper Neutralization of Directives}). 
Notably, these CWE categories are not included in the RAG database of \sys, challenging its ability to repair previously unseen vulnerability types and generalization capability.


\subsection{Experimental Setup and Evaluation Metrics}

\noindent\textbf{LLMs:} 
We employ recent open-source state of the art LLMs as core components within \sys to ensure both effectiveness and practical deployability. 
Specifically, we use \textit{Gemma-4-26B}~\cite{gemma4_2026} for \textsf{Patch Generator} and \textsf{Patch Reviewer} modules, and \textit{Nemotron-3-30B}~\cite{nemotron_30b} for the \textsf{Patch Validator} module. 
To ensure efficient local deployment, all models are executed in 4-bit quantized format using the \texttt{Q4\_K\_M} scheme. Here, \texttt{Q4} denotes 4-bit weight quantization, \texttt{K} refers to a grouped (K-quant) quantization strategy that preserves accuracy through shared scaling factors, and \texttt{M} represents a medium configuration that balances computational efficiency and model fidelity.

\noindent\textbf{Hardware:}
All experiments were conducted on a local AI workstation running Windows 11 Pro, equipped with an Intel® Core™ Ultra 7 processor (up to 5.5 GHz, 20 cores), 32 GB DDR5 memory, and an NVIDIA RTX™ 2000 Ada Generation GPU with 16 GB VRAM. 
The system includes 1 TB PCIe Gen5 NVMe SSD storage and integrated Intel® graphics support.

\noindent\textbf{Evaluation Metrics:} To comprehensively assess the performance of \sys, we evaluate the quality and effectiveness of the generated patches based on three outcome categories: \textit{Correct}, \textit{Partial Fix}, and \textit{Failed}. 

\noindent$\bullet$ \textbf{Correct Fix} implies that the generated patch completely mitigates the underlying vulnerability without introducing new vulnerability and preserves the intended functionality. The patch does not need to match the ground truth exactly at the syntactic or semantic level, provided that it is security-equivalent.

\noindent$\bullet$ \textbf{Partial Fix} indicates that the patch mitigates the security vulnerability but introduces functional inconsistencies or breaks intended functionality. Such patches are still considered partial security fixes, even if they are not fully functionally correct.

\noindent$\bullet$ \textbf{Failed} indicates that the patch does not resolve the underlying vulnerability and the vulnerable behavior remains present. This category also includes patches that introduce new vulnerabilities or other significant issues.

\noindent$\bullet$ \textbf{Repair Success Rate (RSR)} measures the proportion of vulnerabilities successfully addressed by \sys{}, including both fully correct and partial fixes. 
\[
\text{RSR} = (\text{Correct} + \text{Partial Fix}) / \text{Total Instances} \times 100
\]
Following common evaluation practices in prior automated vulnerability repair research \cite{ExtractFix, san2patch, patchagent}, partial fixes are also considered successful repairs when the generated patch mitigates the underlying vulnerability, even if complete functional correctness is not preserved.

\noindent$\bullet$ \textbf{CodeBLEU Score\cite{codebleu}:} a code similarity metric that captures syntactic structure, data-flow, and semantic alignment between the generated patch and the ground truth patch. The score ranges from 0 to 1, where higher values indicate greater similarity.

In our context, CodeBLEU serves as a complementary metric to correctness-based evaluation. A high CodeBLEU score (e.g., $>0.7$) generally indicates strong structural and semantic similarity between the generated patch and the ground truth implementation. However, lower scores do not necessarily imply an incorrect repair, as security-equivalent patches may differ syntactically or adopt alternative implementation strategies while still effectively mitigating the vulnerability.

\subsection{Performance on 10 Java-Based CWEs}

\begin{table*}[t]
\centering
\setlength{\tabcolsep}{3.2pt}
\renewcommand{\arraystretch}{1.11}
\caption{Repair performance for each CWE type. Green denotes correct, olive denotes partial, and red denotes failed fixes}
\label{tab:cwe_results}
\begin{tabularx}{\textwidth}{lrrrrccX}
\toprule
\rowcolor{gray!12}
\textbf{CWE} &
\textbf{Files} &
\textbf{Correct} &
\textbf{Partial} &
\textbf{Failed} &
\makecell{\textbf{Repair}\\\textbf{Success Rate}} &
\makecell{\textbf{CodeBLEU}\\\textbf{Score}} &
\textbf{CVEs} \\
\midrule

CWE-787 & 1 & 0 & 1 & 0 & 100.00\% & 0.630 &
\cvep{CVE-2023-26470 (P)} \\

\midrule

CWE-125 & 3 & 3 & 0 & 0 & 100.00\% & 0.896 &
\cvec{CVE-2020-25021}, \cvec{CVE-2020-25022}, \cvec{CVE-2020-25023} \\

\midrule

CWE-79 & 15 & 9 & 4 & 2 & 86.67\% & 0.872 &
\cvec{CVE-2013-7250}, \cvec{CVE-2014-2065}, \cvep{CVE-2018-1000129 (P)}, \cvec{CVE-2018-25084}, \cvef{CVE-2021-4284 (F)}, \cvec{CVE-2021-43288}, \cvef{CVE-2022-3127 (F)}, \cvec{CVE-2022-32065}, \cvep{CVE-2022-4513 (P)}, \cvec{CVE-2022-4560}, \cvep{CVE-2022-4593 (P)}, \cvec{CVE-2023-29528}, \cvep{CVE-2023-32070 (P)}, \cvec{CVE-2023-33962}, \cvec{CVE-2023-7171} \\

\midrule

CWE-89 & 15 & 7 & 6 & 2 & 86.67\% & 0.772 &
\cvep{CVE-2013-10019 (P)}, \cvep{CVE-2014-125047 (P)}, \cvec{CVE-2014-125052}, \cvep{CVE-2014-125074 (P)}, \cvec{CVE-2015-10020}, \cvep{CVE-2015-10034 (P)}, \cvec{CVE-2016-15021}, \cvec{CVE-2016-4040}, \cvef{CVE-2016-6652 (F)}, \cvep{CVE-2022-4963 (P)}, \cvec{CVE-2023-25157}, \cvep{CVE-2023-25158 (P)} \\

\midrule

CWE-94 & 6 & 1 & 3 & 2 & 66.67\% & 0.844 &
\cvep{CVE-2021-21244 (P)}, \cvec{CVE-2021-21248}, \cvep{CVE-2021-32621 (P)}, \cvep{CVE-2022-24816 (P)}, \cvef{CVE-2022-46166 (F)}, \cvef{CVE-2023-46243 (F)} \\

\midrule

CWE-22 & 21 & 15 & 3 & 3 & 85.71\% & 0.802 &
\cvec{CVE-2018-14371}, \cvef{CVE-2018-9159 (F)}, \cvec{CVE-2020-6950}, \cvec{CVE-2021-32769}, \cvep{CVE-2021-3856 (P)}, \cvec{CVE-2021-39180}, \cvef{CVE-2021-41152 (F)}, \cvep{CVE-2021-41242 (P)}, \cvep{CVE-2021-43795 (P)}, \cvec{CVE-2022-21675}, \cvec{CVE-2022-24830}, \cvec{CVE-2022-29253}, \cvec{CVE-2023-37913}, \cvec{CVE-2024-21633}, \cvef{CVE-2024-24565 (F)} \\

\midrule

CWE-502 & 6 & 3 & 3 & 0 & 100.00\% & 0.879 &
\cvec{CVE-2018-6331}, \cvec{CVE-2021-21242}, \cvep{CVE-2021-21243 (P)}, \cvep{CVE-2021-21249 (P)}, \cvec{CVE-2021-32634}, \cvep{CVE-2022-41958 (P)} \\

\midrule

CWE-352 & 4 & 2 & 1 & 1 & 75.00\% & 0.748 &
\cvec{CVE-2013-7251}, \cvef{CVE-2014-0120 (F)}, \cvep{CVE-2014-0168 (P)} \\

\midrule

CWE-862 & 6 & 3 & 3 & 0 & 100.00\% & 0.849 &
\cvec{CVE-2020-10194}, \cvec{CVE-2022-23617}, \cvep{CVE-2022-23621 (P)}, \cvep{CVE-2022-41929 (P)}, \cvep{CVE-2023-37910 (P)} \\

\midrule

CWE-863 & 6 & 4 & 1 & 1 & 83.33\% & 0.748 &
\cvep{CVE-2020-19005 (P)}, \cvec{CVE-2021-21318}, \cvec{CVE-2021-32620}, \cvec{CVE-2022-39302}, \cvef{CVE-2023-26056 (F)} \\

\midrule
\rowcolor{green!8}
\textbf{10 CWEs} & \textbf{83} & \textbf{47} & \textbf{25} & \textbf{11} & \textbf{86.75\%} & \textbf{0.815} &
\textbf{--} \\

\bottomrule
\end{tabularx}
\end{table*}

The results presented in \Cref{tab:cwe_results} demonstrate the effectiveness of \sys{} across diverse vulnerability families. For \textit{memory-based vulnerabilities} (CWE-787, CWE-125), \sys{} achieves a perfect repair success rate of \textbf{100\%}, with all instances successfully mitigated, indicating strong capability in handling localized memory safety issues. Moving to \textit{injection-based vulnerabilities} (CWE-79, CWE-89, CWE-94), the framework achieves high success rates of \textbf{86.67\%} for both XSS and SQL injection, while code injection (CWE-94) achieves a success rate of \textbf{66.67\%}, reflecting the increased complexity of semantic program transformations required for such vulnerabilities. For \textit{input handling vulnerabilities} (CWE-22, CWE-502), \sys{} maintains strong performance with success rates of \textbf{85.71\%} and \textbf{100.00\%}, respectively, demonstrating its ability to handle vulnerabilities involving external input validation and sanitization across different contexts.

\textit{Access control vulnerabilities} (CWE-352, CWE-862, CWE-863) exhibit consistently strong but still varied performance, with success rates ranging from \textbf{75.00\%} to \textbf{100.00\%}. While \sys{} effectively mitigates all cases of missing authorization (CWE-862 at \textbf{100.00\%}), the relatively lower performance on CWE-352 highlights the challenges associated with correctly identifying and enforcing security checks in access-control-related workflows. Overall, out of \textbf{83} real-world vulnerabilities, \sys{} successfully repairs \textbf{72} instances (including both correct and partial fixes), achieving an overall repair success rate of \textbf{86.75\%} across all $10$ CWE categories, indicating robust performance across diverse vulnerability types.
Furthermore, the consistently high CodeBLEU scores (average \textbf{0.815}) suggest that the generated patches not only address the vulnerabilities but also maintain strong syntactic and semantic similarity to the ground truth, underlying the quality and reliability of the repair process.
To the best of our knowledge, no existing automated repair framework explicitly targets Java-based CWEs spanning \textit{access control, SQL injection, code injection, and memory-related vulnerabilities}. As a result, direct benchmark comparisons are currently unavailable, and the results reported for \sys{} provide an initial reference point for this setting.

\begin{table*}[t]
\centering
\normalsize
\setlength{\tabcolsep}{4pt}
\renewcommand{\arraystretch}{1.15}
\caption{Comparison with recent existing repair tools. Baseline results are reported from their papers; \cmark denote correct and partial repairs; \xmark  denote failures, and ``--'' denotes unavailable}
\label{tab:benchmark_comparison}
\begin{adjustbox}{max width=\textwidth}
\begin{tabular}{lllcccccc}
\toprule
\rowcolor{gray!12}
\textbf{CVE} & \textbf{CWE} & \textbf{Target File} & \textbf{\sys} & \textbf{SAN2PATCH}\cite{san2patch} & \textbf{PatchAgent}\cite{patchagent} & \textbf{ExtractFix}\cite{ExtractFix} & \textbf{VulnFix}\cite{VulnFix} & \textbf{VulMaster}\cite{Vulmaster} \\
\midrule

\rowcolor{gray!8}
CVE-2016-8691  & CWE-369 & \texttt{jpc\_cs.c} & \cmark & \xmark & \cmark & \cmark & \cmark & \cmark \\

CVE-2016-9387  & CWE-190 & \texttt{jpc\_dec.c} & \cmark & -- & \cmark & \cmark & -- & -- \\

\rowcolor{gray!8}
CVE-2018-14498 & CWE-125 & \texttt{rdbmp.c} & \cmark & \cmark & \cmark & \xmark & \xmark & \xmark \\

CVE-2016-5314  & CWE-787 & \texttt{tif\_pixarlog.c} & \xmark & -- & \cmark & \xmark & -- & -- \\

\rowcolor{gray!8}
CVE-2016-9557  & CWE-190 & \texttt{jpg\_dec.c} & \cmark & \cmark & -- & -- & \xmark & \xmark \\

CVE-2016-5844  & CWE-190 & \texttt{archive\_read\_support\_format\_iso9660.c} & \cmark & \cmark & -- & -- & \cmark & \xmark \\

\rowcolor{gray!8}
CVE-2016-10092 & CWE-119 & \texttt{tiffcrop.c} & \cmark & \cmark & -- & -- & \xmark & \xmark \\

CVE-2016-10272 & CWE-119 & \texttt{tiffcrop.c} & \cmark & \cmark & -- & -- & \xmark & \xmark \\

\rowcolor{gray!8}
CVE-2017-5225  & CWE-119 & \texttt{tiffcp.c} & \xmark & \cmark & -- & -- & \xmark & \xmark \\

\bottomrule
\end{tabular}
\end{adjustbox}
\vspace{-0.6em}
\end{table*}

\subsection{Generalizability on Unseen Vulnerabilities}
\label{sec:generalizability} To evaluate the generalization capability of \sys{} and enable comparison with existing repair tools, we consider a set of real-world CVEs from the CVEFixes dataset~\cite{CVEFixes} that fall outside our targeted CWE categories but are shared across multiple prior repair frameworks. These vulnerabilities are in C programs, which introduces additional challenge such as cross-language repair with the limited availability of closely related vulnerability patches within the retrieval database. As shown in \Cref{tab:benchmark_comparison}, the selected CVEs cover a range of memory safety and arithmetic-related weaknesses, including CWE-119, CWE-125, CWE-190, CWE-369, and CWE-787.

Across these instances, \sys{} produces correct fixes for \textbf{7 out of 9} CVEs, failing on two cases corresponding to CWE-787 and CWE-119. For comparison, we report baseline results as presented in the respective papers, while \sys{} is evaluated directly on the selected CVEs. At the per-CVE level, SAN2PATCH~\cite{san2patch} successfully repairs several memory-related vulnerabilities (e.g., CWE-125 and CWE-119), but fails or is not applicable in other cases such as CWE-369. PatchAgent~\cite{patchagent} exhibits complementary behavior, successfully handling some arithmetic-related vulnerabilities (e.g., CWE-190 and CWE-369), but without consistent coverage across all evaluated instances. In contrast, \sys{} demonstrates competitive performance across most CVEs despite not being specifically designed for these vulnerability categories. Other approaches, including ExtractFix~\cite{ExtractFix}, VulnFix~\cite{VulnFix}, and VulMaster~\cite{Vulmaster}, show limited or inconsistent coverage across the evaluated CVEs, with multiple instances either unsupported or failing to produce valid repairs. These results highlight the difficulty of generalizing automated vulnerability repair across diverse vulnerability classes and suggest that \sys{} generalizes effectively to previously unseen vulnerabilities.

\subsection{Evaluation on Out-of-Distribution data}

\subsubsection{Generalization on Vul4J Dataset}

To further evaluate the generalization capability of \sys, we conduct experiment on a subset of \textbf{23} randomly selected CVEs from the Vul4J\cite{vul4j2022} dataset, specifically focusing on the \textit{single-file-commit} setting. 
These vulnerabilities span multiple CWE categories (e.g., CWE-835, CWE-20, CWE-22, CWE-611, CWE-79, CWE-269, CWE-310, CWE-532, CWE-770, CWE-918), most of which are not part of the targeted CWE categories in our retrieval database, thereby representing a true out-of-distribution evaluation setting. 
As shown in \Cref{tab:ood_results}, \sys produces \textbf{15} successful repairs including \textbf{10} correct and \textbf{5} partial fixes, resulting in a repair success rate of \textbf{65.22\%}.
Despite the absence of similar vulnerability patterns in the retrieval database and the diversity of unseen CWE types, the framework demonstrates remarkable performance across these CVEs. Furthermore, the average CodeBLEU score of \textbf{0.897} indicates that the generated patches maintain strong structural and semantic similarity to the ground truth.
Overall, these results suggest that \sys{} is capable of adapting to unseen vulnerability types and generating meaningful repairs beyond its primary evaluation scope.

\begin{table}[t]
\centering
\setlength{\tabcolsep}{2.8pt}
\renewcommand{\arraystretch}{1.13}
\caption{\sys{} performance on out-of-distribution data.}
\label{tab:ood_results}

\begin{tabular}{|p{0.29\columnwidth}|c|c|c|c|c|c|}
\hline
\rowcolor{gray!12}
\textbf{CWEs} &
\textbf{CVEs} &
\textbf{Correct} &
\textbf{Partial} &
\textbf{Fail} &
\makecell{\textbf{Succ.}\\\textbf{Rate}} &
\makecell{\textbf{Code}\\\textbf{BLEU}} \\
\hline

\makecell[l]{CWE-835, CWE-20,\\ 
CWE-22, CWE-611,\\
CWE-79, CWE-269,\\ 
CWE-310, CWE-532,\\
CWE-770, CWE-918} &
23 & 10 & 5 & 8 & 65.22\% & 0.897 \\
\hline

\end{tabular}
\vspace{-0.6em}
\end{table}

\subsubsection{Evaluation on Recent Vulnerabilities}

Furthermore, we evaluate \sys on a subset of vulnerabilities from the \textit{Zero-Day} dataset introduced in APPATCH~\cite{appatch}, consisting of recently disclosed CVEs (2024) in C programs. These instances cover diverse CWE categories (e.g., CWE-835, CWE-828, CWE-770, CWE-476, CWE-416, CWE-22, CWE-281, CWE-96, CWE-125), including some vulnerability types that are not part of the targeted CWE categories in our retrieval database.
As shown in \Cref{tab:zeroday_results}, \sys{} successfully repairs \textbf{14 out of 16} cases, including \textbf{11} correct and \textbf{3} partial fixes, with only two failed instances. 
We note that APPATCH\cite{appatch} does not report per-CVE repair results, which limits direct comparison at a fine-grained level. 
Nevertheless, the observed results indicate that \sys is capable of generating effective repairs on recently disclosed and diverse vulnerability instances.

\begin{table}[t]
\centering
\setlength{\tabcolsep}{4pt}
\renewcommand{\arraystretch}{1.15}
\caption{\sys{} performance on 2024 CVEs from APPATCH\cite{appatch} dataset. \cmark mark correct and partial repairs; \xmark  mark failures}
\label{tab:zeroday_results}

\begin{tabular}{|l|c|p{2.2cm}|c|}
\hline
\rowcolor{gray!12}
\textbf{CVE} & \textbf{CWE} & \textbf{Target File} & \textbf{\sys} \\
\hline

\rowcolor{gray!8}
CVE-2024-2397  & CWE-835 & \texttt{print-ppp.c} & \xmark \\
CVE-2024-2397  & CWE-835 & \texttt{print.c} & \cmark \\

\rowcolor{gray!8}
CVE-2024-24746 & CWE-835 & \texttt{ble\_hs\_conn.c} & \cmark \\
CVE-2024-25742 & CWE-828 & \texttt{sev-shared.c} & \cmark \\

\rowcolor{gray!8}
CVE-2024-25742 & CWE-828 & \texttt{sev.c} & \cmark \\
CVE-2024-28871 & CWE-770 & \texttt{http\_request.c} & \cmark \\

\rowcolor{gray!8}
CVE-2024-29489 & CWE-476 & \texttt{ecma-function-object.c} & \cmark \\
CVE-2024-29489 & CWE-476 & \texttt{ecma-proxy-object.c} & \cmark \\

\rowcolor{gray!8}
CVE-2024-31578 & CWE-416 & \texttt{hwcontext.c} & \cmark \\
CVE-2024-32002 & CWE-22  & \texttt{submodule--helper.c} & \cmark \\

\rowcolor{gray!8}
CVE-2024-32020 & CWE-281 & \texttt{clone.c} & \cmark \\
CVE-2024-32487 & CWE-96  & \texttt{filename.c} & \cmark \\

\rowcolor{gray!8}
CVE-2024-32658 & CWE-125 & \texttt{interleaved.c} & \cmark \\
CVE-2024-32659 & CWE-125 & \texttt{color.c} & \cmark \\

\rowcolor{gray!8}
CVE-2024-32661 & CWE-476 & \texttt{info.c} & \xmark \\
CVE-2024-32662 & CWE-125 & \texttt{redirection.c} & \cmark \\

\hline
\end{tabular}
\vspace{-0.6em}
\end{table}

\subsection{\sys's Efficiency}

\begin{table*}[!t]
\centering
\setlength{\tabcolsep}{5pt}
\renewcommand{\arraystretch}{1.15}
\caption{\sys efficiency in terms of token and time cost}
\label{tab:raven_efficiency}
\begin{tabular}{lcccc}
\toprule
\rowcolor{gray!12}
\textbf{Vulnerabilities} & 
\makecell{\textbf{Semantic Retrieval}\\\textbf{(tokens / sec)}} & 
\makecell{\textbf{Context Retrieval}\\\textbf{(tokens / sec)}} & 
\makecell{\textbf{Patch Generator}\\\textbf{(tokens / sec)}} & 
\makecell{\textbf{Patch Reviewer}\\\textbf{(tokens / sec)}} \\
\midrule

\Cref{tab:cwe_results} Average on 10 Java CWEs & 13013 / 48.15 & 100108 / 147 & 30469 / 172.25 & 14282 / 50.4 \\

\Cref{tab:cwe_results_c_lang} C CWEs (125 and 863) & 12943 / 70.2 & 86379 / 194.5 & 35981 / 253.4 & 17179 / 78.4 \\

\Cref{tab:benchmark_comparison} Unseen Vulnerabilities in C & 40745 / 64.3 & 142487 / 155.9 & 161009 / 261.8 & 147343 / 118.3 \\

\Cref{tab:zeroday_results} Unseen CVEs 2024 from APPATCH & 14560 / 38.7 & 116208 / 148.2 & 37385 / 196.7 & 23987 / 52.2 \\

\midrule
\rowcolor{green!8}
\textbf{Average} & \textbf{14404 / 51.9} & \textbf{99509 / 155.0} & \textbf{41773 / 206.4} & \textbf{24487 / 62.4} \\

\bottomrule
\end{tabular}
\vspace{-0.6em}
\end{table*}

\noindent\textbf{Token Cost} 
\sys{} leverages the Google Gemma-4-26B~\cite{gemma4_2026} model in a fully local deployment setting. As an open-source model, it eliminates reliance on proprietary LLM services and associated API costs. As shown in \Cref{tab:raven_efficiency}, the \textsf{Context Retrieval} module accounts for the highest token consumption across all evaluation settings (e.g., \textbf{99,509} tokens on average), since it is responsible for repository-level analysis, including the retrieval of cross-file dependencies and global contextual information.

In contrast, the \textsf{Semantic Retriever} and \textsf{Patch Reviewer} modules exhibit comparatively lower token usage (e.g., \textbf{14,404} and \textbf{24,487} tokens on average, respectively), indicating a more balanced distribution of computational cost across the pipeline.

\noindent\textbf{Time Cost}
In terms of execution time, \sys remains efficient despite operating in a local setup. 
The \textsf{Context Retrieval} and \textsf{Patch Generator} modules account for the majority of the runtime, with average times of \textbf{155.0} seconds and \textbf{206.4} seconds across all the experimental settings as shown in~\Cref{tab:raven_efficiency}. 
The higher cost in the \textsf{Patch Generator} module is attributed to the automated iterative repair process, where patches are refined over multiple controlled iterations. 
Although this iterative process increases execution time, it significantly improves repair quality, particularly for complex vulnerabilities. 
Overall, the end-to-end repair time of \sys{} averages around \textbf{7--8 minutes per vulnerability}, which is substantially lower than recent tools such as PatchAgent~\cite{patchagent}, where even the fastest configuration reports \textbf{28.7 minutes} per repair, demonstrating a significant improvement in repair efficiency. 

Overall, the experimental results demonstrate that \sys{} can repair a broad range of real-world vulnerabilities across diverse CWE categories and evaluation settings. Beyond strong performance on the targeted Java-based vulnerabilities, \sys{} also generalizes to previously unseen vulnerability types, recent CVEs, cross-dataset evaluations, and vulnerabilities in C programs. The results further suggest that the combination of agentic retrieval, repository-level contextual reasoning, and iterative refinement enables \sys{} to generate effective and semantically meaningful repairs across diverse vulnerability classes. Consistently high CodeBLEU scores additionally indicate strong structural and semantic alignment with ground truth fixes, highlighting the potential of agentic RAG-based approaches for scalable and context-aware automated vulnerability repair.

\section{Related Work}
\label{sec:related_work}
\noindent\textbf{Automated Program Repair (APR)}\label{sec:related_work:tee_migration}
\textit{Search-based APR} aims to identify correct patches from a predefined search space using mutation operators and test-suite validation. Approaches such as CapGen\cite{CAPGEN}, sharpFix\cite{sharpFix}, and VarFix~\cite{VarFix} leverage contextual information, code reuse, and variational execution but they often suffer from search space explosion and frequently fail when the correct patch is absent from the mutation space~\cite{mutation}. 
\textit{Semantic-based APR} reduces the search space by leveraging symbolic execution and program semantics to derive repair constraints~\cite{repairconstraints}. Techniques such as SOSRepair\cite{SOSrepair}, Maple\cite{maple}, VulnFix~\cite{VulnFix} and ExtractFix~\cite{ExtractFix} generate patches that satisfy these constraints, improving efficiency but limiting adaptability when formal constraints fail to capture the underlying bug semantics.

\textit{Template-based APR} relies on predefined fix patterns extracted either manually or automatically. 
Works such as PAR~\cite{PAR}, kPAR~\cite{KPAR}, and Relifix~\cite{relifix} use expert-defined patterns, while Genesis~\cite{Genesis}, AVATAR~\cite{AVATAR}, and TBar~\cite{Tbar} automate pattern generation. 
While effective for recurring bug patterns, these methods depend heavily on accurate fault localization and suitable templates, restricting generalization to unseen or complex vulnerabilities.

\textit{Learning-based APR}
models program repair as a sequence-to-sequence transformation from buggy to fixed code.
Approaches such as Sequencer~\cite{sequence} and DeepRepair~\cite{deeprepair} use neural models and code similarity to improve repair generation, while DLFix~\cite{dlfix}, CoCoNuT~\cite{10.1145/3395363.3397369}, and Review4Repair~\cite{review4repair} incorporate program representations (ASTs) or human-written review information for richer context. 
Nevertheless, these methods remain limited for complex, context-dependent vulnerabilities due to their reliance on localized representations and training data.

\noindent\textbf{LLM-based Vulnerability Repair} \label{sec:related_work:rnn}PatchAgent~\cite{patchagent} introduces an agent-based framework combining fault localization, patch generation, and validation. However, it requires substantial repair time and incurs high financial costs due to its reliance on proprietary LLMs. APPATCH~\cite{appatch} proposes an LLM-based framework using adaptive prompting. Nevertheless, it depends on explicitly providing vulnerability location and type within prompts, limiting its autonomy. SAN2PATCH~\cite{san2patch} leverages sanitizer logs with multi-stage LLM reasoning to generate patches. However, its applicability is limited to vulnerabilities with available sanitizer logs. Moreover, these frameworks~\cite{san2patch,appatch,patchagent} focus only on memory and arithmetic-related vulnerabilities. VulMaster~\cite{Vulmaster} employs a CodeT5-based model incorporating AST structure and CWE knowledge for patch generation. However, it relies on explicitly specified modification locations. Pearce \etal~\cite{zeroshotrepair} study zero-shot vulnerability repair, highlighting prompt-design challenges in generating suitable fixes. Although effective on synthetic cases, it struggles to produce functionally correct patches for real CVEs. RAP-Gen~\cite{rapgen} introduces a retrieval-augmented LLM framework that leverages CodeT5 to retrieve relevant bug-fix patterns through lexical and semantic matching. Zhang \etal~\cite{Javaeffective} present a comparative study evaluating LLMs and DL-based APR frameworks. VRepair~\cite{vrepair} and SeqTrans~\cite{seqtrans} leverage pre-training on large-scale bug-fixing data followed by fine-tuning on smaller vulnerability datasets. Nevertheless, VulRepair~\cite{Vulrepair} outperforms both approaches, although some of its limitations are later addressed by VulMaster~\cite{Vulmaster}.

\noindent\textbf{Summary}   
It is evident that, although various frameworks have been proposed, none match the effectiveness of \sys{} in achieving high repair success across different experimental settings. \sys{} is the first framework to target diverse vulnerability types and multiple programming languages while demonstrating strong generalization to previously unseen vulnerability types with negligible repair cost.

\section{Conclusion}
\label{sec:conclusion}

To conclude, \sys{} presents a scalable and autonomous framework for automated vulnerability repair based on agentic retrieval-augmented generation and an iterative repair mechanism. By combining historical fix retrieval, repository-level context retrieval, and iterative patch refinement, \sys{} effectively addresses complex vulnerabilities that cannot be resolved using local vulnerable code alone. Our evaluation demonstrates strong repair performance across diverse vulnerability types, unseen CWE categories, and additional programming language settings, while maintaining efficient, low-cost deployment through locally deployable open-source LLMs.

\bibliographystyle{IEEEtran}
\bibliography{bib}
\appendix
\label{sec:appendix}
This appendix provides background on Retrieval Augmented Generation, prompt templates and  reports the results of repair performance on C based CWEs. 

\begin{table*}[t]
\centering
\scriptsize
\setlength{\tabcolsep}{3.5pt}
\renewcommand{\arraystretch}{1.18}
\caption{Repair performance for C CWEs. Green denotes correct, olive denotes partial, and red denotes failed fix}
\label{tab:cwe_results_c_lang}
\begin{tabularx}{\textwidth}{lrrrrc c X}
\toprule
\rowcolor{gray!12}
\textbf{CWE} &
\textbf{Files} &
\textbf{Correct} &
\textbf{Partial} &
\textbf{Failed} &
\makecell{\textbf{Repair}\\\textbf{Success Rate}} &
\makecell{\textbf{CodeBLEU}\\\textbf{Score}} &
\textbf{CVEs} \\
\midrule

CWE-125 & 20 & 15 & 2 & 3 & 85.00\% & 0.894 &
\cvec{CVE-2015-2697}, \cvec{CVE-2016-6214}, \cvef{CVE-2016-6911 (F)}, \cvec{CVE-2017-6309}, \cvec{CVE-2017-12897}, \cvec{CVE-2017-12985}, \cvec{CVE-2017-13007}, \cvec{CVE-2017-13014}, \cvef{CVE-2017-13045 (F)}, \cvec{CVE-2017-17081}, \cvec{CVE-2018-20174}, \cvec{CVE-2018-20854}, \cvec{CVE-2019-19274}, \cvec{CVE-2020-4033}, \cvep{CVE-2020-11089 (P)}, \cvec{CVE-2020-15888}, \cvep{CVE-2021-3881 (P)}, \cvef{CVE-2022-23467 (F)} \\

\midrule

CWE-863 & 9 & 3 & 5 & 1 & 88.89\% & 0.754 &
\cvec{CVE-2018-18397}, \cvep{CVE-2018-18955 (P)}, \cvec{CVE-2018-20685}, \cvec{CVE-2019-15900}, \cvep{CVE-2020-24716 (P)}, \cvep{CVE-2023-5521 (P)}, \cvef{CVE-2023-5521 (F)}, \cvep{CVE-2023-46139 (P)} \\

\midrule
\rowcolor{green!8}
\textbf{2 CWEs} & \textbf{29} & \textbf{18} & \textbf{7} & \textbf{4} & \textbf{86.94\%} & \textbf{0.824} &
\textbf{--} \\

\bottomrule
\end{tabularx}
\end{table*}

\subsection{Retrieval Augmented Generation}
\label{sec:background}
Retrieval Augmented Generation (RAG) incorporates external knowledge at inference time by combining the internal knowledge of an LLM with external sources such as vector databases, knowledge bases, and code repositories~\cite{lewis2020retrieval,gao2024rag}. This is particularly beneficial for knowledge-intensive tasks where information changes dynamically or where expert knowledge is difficult to acquire reliably during pre-training. 

Compared with fine-tuning, RAG offers greater flexibility and scalability, as external knowledge can be updated independently of the model parameters, reducing the need for costly retraining. It also improves factual grounding and transparency by enabling the model to generate responses based on retrieved, context-specific evidence, thereby mitigating hallucinations and enhancing trustworthiness in downstream applications.

A typical RAG workflow begins by constructing an external knowledge base, where raw data are collected, processed into retrievable documents, and enriched with relevant metadata. At inference time, the most relevant documents are selected based on the query using lexical or semantic retrieval. During response generation, the query and retrieved context are incorporated into a prompt template, guiding the LLM and substantially improving output quality. However, irrelevant or conflicting retrieved documents can mislead the model, making retrieval quality and prompt design critical for generating reliable, high-quality outputs.

An emerging extension of a RAG paradigm is Agentic RAG, which augments standard retrieval-generation pipelines with autonomous reasoning and decision-making capabilities. Instead of performing retrieval only once, an agentic system can iteratively plan retrieval steps, reformulate queries, evaluate intermediate results, and selectively invoke external tools or additional knowledge sources as needed. This enables dynamic adaptation to complex, multi-step tasks where relevant information may not be accessible through a single retrieval pass. By integrating retrieval with iterative reasoning and tool use, Agentic RAG can improve robustness, contextual understanding, and task completion accuracy, particularly in domains that require structured problem solving or continuous interaction with evolving external information.

\subsection{Performance of \sys on C based CWEs}

The results in \cref{tab:cwe_results_c_lang} further demonstrate the effectiveness of \sys on C-based vulnerabilities across memory-safety and access-control categories. 
For \textit{memory-based vulnerabilities} (CWE-125), \sys{} successfully repairs \textbf{17} out of \textbf{20} instances, achieving a repair success rate of \textbf{85.00\%}, which shows its ability to mitigate out-of-bounds read vulnerabilities that often require precise boundary and input-size reasoning. 
For \textit{access-control vulnerabilities} (CWE-863), \sys{} achieves an even higher success rate of \textbf{88.89\%}, successfully repairing \textbf{8} out of \textbf{9} instances, indicating that the framework can also handle security flaws requiring authorization-aware reasoning beyond local code patterns. 
Overall, across \textbf{29} C vulnerabilities, \sys repairs \textbf{25} instances including correct and partial fixes, achieving an overall repair success rate of \textbf{86.94\%}, while the average CodeBLEU score of \textbf{0.824} highlights that the generated patches remain closely aligned with the ground-truth repairs. 

\subsection{Prompt Templates} \label{appendix:templates}

\begin{figure*}[t]
\centering
\begin{tcolorbox}[
    colback=gray!5!white,
    colframe=black,
    width=\textwidth,
    boxrule=0.5pt,
    arc=4pt,
    outer arc=4pt
]

\textbf{Selector Agent Prompt Template:}

\begin{lstlisting}[basicstyle=\ttfamily\small, breaklines=true]
You are a <language> code vulnerability patch selection agent.

Select the single candidate patch whose fix strategy is the best semantic reference for repairing the vulnerable code.

You are given:
- a vulnerable code snippet
- a CWE description
- several candidate patches

Your task is to choose the candidate whose remediation approach is most similar to the fix that should be applied to the vulnerable code.

Evaluation rules:
1. Prioritize the underlying vulnerability cause over file names, function names, or CVE IDs.
2. Prefer the candidate with the most similar mitigation strategy and security checks.
3. Focus on the security logic introduced by the patch (for example: bounds validation, length checks, null checks, state validation, safe iteration limits).
4. Do not rely on superficial textual overlap alone.
5. If the CWE description and vulnerable code appear inconsistent, prioritize the actual vulnerable code pattern and the implied root cause in the code.
6. If a candidate appears to already match logic present in the vulnerable code, treat it cautiously; it may indicate dataset noise or a pre-patched snippet.
7. Assume candidate patches may come from related or unrelated files; file identity is not a deciding factor.

Return exactly:
Candidate x

vulnerable_code:
<vulnerable_code>

cwe_description:
<cwe_description>

given_patches:

Candidate 1
CVE: <candidate_1_cve_id>
Filename: <candidate_1_filename>
Patch diff:
<candidate_1_patch_diff>

---

Candidate 2
CVE: <candidate_2_cve_id>
Filename: <candidate_2_filename>
Patch diff:
<candidate_2_patch_diff>

Return exactly:
Candidate x
\end{lstlisting}

\end{tcolorbox}
\caption{Prompt template for Selection Agent}
\label{fig:selector-prompt}
\end{figure*}

\begin{figure*}[t]
\centering
\begin{tcolorbox}[
    colback=gray!5!white,
    colframe=black,
    width=\textwidth,
    boxrule=0.5pt,
    arc=4pt,
    outer arc=4pt
]

\textbf{Context Curator Agent Prompt Template:}

\begin{lstlisting}[basicstyle=\ttfamily\small, breaklines=true]
You are a repository context curator for <language> vulnerability repair.

Your only job is to read the checked-out repository, identify the few most relevant files, and prepare compact evidence for a separate repair actor.

Never produce code patches.
Never use tools outside the allowed list.

The target file `code_before` is already provided below. Use function tools to inspect other repository files, or to narrow supporting regions when needed.

Case summary:
- Case: <case_label>
- CVE/CWE: <cve_id> / <cwe_id>
- Target filename: <target_filename>
- Target path: <target_relpath>
- Current step: <step_index> / <max_steps>
- Invalid JSON responses so far: <invalid_response_count>

Reference patch commit message from the actual vulnerability-fixing commit:
<commit_message_or_unavailable>

Goal:
- Find the smallest set of repository files that explain the vulnerable runtime path.
- Prioritize the target file, direct callers/callees, and closely related validation helpers.
- Finish once you can justify the most relevant file snippets for the downstream actor prompt.
- If you use search alternation like `foo|bar`, set `mode` to `regex`, not `literal`.
- If you do not set `glob` for `search_repo`, the tool will default to <default_source_glob> for this case.

Confirmed facts:
<confirmed_facts_or_none>

Recent trace:
<recent_tool_trace>

Target file code_before:
<vulnerable_code_with_line_numbers>

Available function tools:
- read_bundle()
- read_file(path, start_line, end_line)
- search_repo(pattern, mode, glob, limit)
- find_symbol(name, include_target)
- finish_curating(root_cause, constraints, selected_files)

Tool rules:
- Use OpenAI-style function calls via the provided tools.
- Do not write raw JSON in assistant content to simulate a tool call.
- Finish by calling `finish_curating`, not by answering in plain text.
- `selected_files` must contain at most 5 items.
- Include at least one snippet from the target file.
- The target file `code_before` is already provided in the prompt; use tools mainly to inspect other repository files or to narrow supporting evidence.
- In curator mode, `find_symbol` defaults to `include_target=true` if you do not specify it.
- Lower `priority` means more important.
\end{lstlisting}

\end{tcolorbox}
\caption{Prompt template for context retrieval}
\label{fig:context-curator-prompt}
\end{figure*}

\begin{figure*}[t]
\centering
\begin{tcolorbox}[
    colback=gray!5!white,
    colframe=black,
    width=\textwidth,
    boxrule=0.5pt,
    arc=4pt,
    outer arc=4pt
]
\textbf{Patch Generator Prompt Template:}

\begin{lstlisting}[basicstyle=\ttfamily\small, breaklines=true]
You are a software vulnerability repair tool.

Generate the smallest correct {language} fix for the target file based on the vulnerable code below.

Your goal is to repair the actual vulnerability in the target code, not to imitate the reference patch mechanically.

Rules:
1. Prioritize the concrete root cause visible in the target code.
{reference_guidance}
4. Apply the smallest targeted fix that prevents the unsafe behavior while preserving existing logic.
5. Do not refactor, rename symbols, change formatting unnecessarily, or modify unrelated behavior.
6. Do not introduce new helper functions, macros, types, or dependencies unless clearly implied by the provided code context.
7. Reuse existing validation style, control flow, and error handling patterns when possible.
8. Ensure the result is syntactically valid and consistent with the surrounding code.

Output rules:
- Return only one ```{code_fence}``` code block and no other text.
- Inside the code block, provide only the minimal fixed code snippet(s) for the target file.
- Do not output a unified diff unless explicitly requested.
{reference_output_rule}

Target {source_label}:
{case.target_filename}

vulnerable_code:
{case.buggy_code}

cwe_description:
{case.cwe_description}

{reference_section}

C/C++ requirements:
- Return the smallest plausible C/C++ code replacement snippet(s) for the target file.
- Do not invent classes, wrappers, namespaces, packages, or non-existent project abstractions.
- Use existing variables, macros, and error paths where possible.
- Add checks immediately before the unsafe access or operation they protect.
- Do not introduce undefined labels, variables, or return values inconsistent with the surrounding function.

Optional repository context:
repo_context_curated_from_repository:
{repo_context_prompt}

revision context:
previous_fix:
{previous_fix}

critic_feedback:
{critic_feedback}

Revision requirements:
- Improve the previous fix using the critic feedback.
- Correct the specific mistakes identified by the critic.
- Do not repeat flaws from the previous fix.
- If the previous fix conflicts with the actual vulnerable pattern in the target code, prioritize the target code and the critic feedback.
\end{lstlisting}

\end{tcolorbox}
\caption{Prompt template for Patch Generator}
\label{fig:patch-generator-prompt}
\end{figure*}

\end{document}